\documentclass{article}

 \usepackage[nonanonymous]{neurips_2026}
\nolinenumbers

\usepackage{iftex}
\ifPDFTeX
\usepackage[utf8]{inputenc}
\usepackage[T1]{fontenc}
\else
\usepackage{fontspec}
\usepackage{xeCJK}
\setCJKsansfont{FandolHei-Regular.otf}
\setCJKmonofont{FandolHei-Regular.otf}

\fi
\usepackage{hyperref}
\usepackage{url}
\usepackage{booktabs}
\usepackage{amsfonts}
\usepackage{amssymb}
\usepackage{nicefrac}
\usepackage{microtype}
\usepackage{graphicx}
\usepackage{multirow}
\usepackage{amsmath}
\usepackage{multicol}
\usepackage{xcolor}
\usepackage{cleveref}
\usepackage{subcaption}
\usepackage[normalem]{ulem}
\usepackage{fvextra}
\usepackage{tcolorbox}
\tcbuselibrary{breakable, skins}
\usepackage{listings}

\definecolor{xiaomiblue}{HTML}{4A7BCE}
\definecolor{xiaomipaleblue}{HTML}{B8DCFE}
\definecolor{xiaomiorange}{HTML}{FFA903}
\definecolor{xiaomiteal}{HTML}{03CCA0}
\definecolor{xiaomigreen}{HTML}{50B341}
\definecolor{xiaomicoral}{HTML}{ED696D}
\definecolor{xiaomilightgray}{HTML}{AAAAA8}
\definecolor{xiaomibrightblue}{HTML}{04A3FD}
\definecolor{xiaomimedgray}{HTML}{6E6E6C}
\definecolor{xiaomiblack}{HTML}{030303}
\definecolor{xiaomired}{HTML}{ee4028}
\definecolor{salmon}{HTML}{FFA07A}
\definecolor{highlightblue}{HTML}{E1F5FE}
\definecolor{baselinegray}{HTML}{F2F2F2}

\newcommand{\cmark}{\textcolor{green!60!black}{\checkmark}}
\newcommand{\xmark}{\textcolor{red}{$\times$}}
\newcommand{\naall}{\textcolor{gray!60}{-}}

\newcommand{\best}[1]{\textbf{#1}}
\newcommand{\second}[1]{\underline{#1}}

\title{\textcolor{xiaomiorange}{Mi}DashengLM-Gen: Unified Audio Scene Generation via LLM-Driven Autoregressive Flow Matching}

\author{%
  Xingwei Sun$^{1}$
  \quad Heinrich Dinkel$^{1}$
  \quad Gang Li$^{1}$  
  \quad Jiahao Mei$^{1,2}$  
  \quad Yadong Niu$^{1}$
  \\ \vspace{0.2cm} \\
  \quad \textbf{Zerui Han}$^{1}$
  \quad \textbf{Yuepeng Jiang}$^{1}$
  \quad \textbf{Jiahao Zhou}$^{1}$ 
  \quad \textbf{Lichun Fan}$^{1}$  
  \quad \textbf{Jian Luan}$^{1}$ \quad\\
  \vspace{0.2cm} \\
  $^{1}$MiLM Plus, Xiaomi Inc., Beijing, China\\  
  $^{2}$X-LANCE Lab, Shanghai Jiao Tong University, Shanghai, China\\
}
 
\begin{document}

\maketitle

\begin{abstract}
Generating coherent audio scenes that simultaneously blend speech, music, and sound effects remains a significant challenge.
Current approaches typically rely on a disjointed pipeline where a frozen, decoupled text encoder feeds a separate audio decoder, limiting cross-modal optimization and leading to poor speech intelligibility.
To overcome these limitations, we introduce MiDashengLM-Gen, an end-to-end framework that couples a pre-trained Large Language Model (LLM) with per-token conditional flow matching for autoregressive, variable-length mixed-audio scene generation.
MiDashengLM-Gen represents a first approach for general text-to-audio generation with one end-to-end trained model.
Empirical evaluations demonstrate that MiDashengLM-Gen drastically improves speech intelligibility over existing unified models.
On the Seed-TTS benchmark, English Word Error Rate (WER) drops from 12.15\% to 2.79\%, approaching the performance of dedicated Text-to-Speech (TTS) systems (1.24\%). 
Furthermore, the framework extends effectively to multilingual settings, yielding highly competitive multilingual WERs compared to existing baselines.
Lastly, the model maintains competitive mixed-audio generation quality on the MECAT benchmark. Code and checkpoints are available at \href{https://github.com/xiaomi-research/midashenglm-gen}{\includegraphics[height=1em,keepaspectratio]{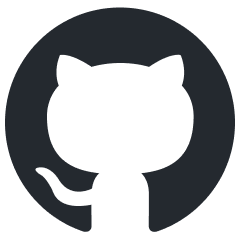}}\footnote{\url{https://github.com/xiaomi-research/midashenglm-gen}} and \href{https://huggingface.co/mispeech/midashenglm-gen}{\includegraphics[height=1em,keepaspectratio]{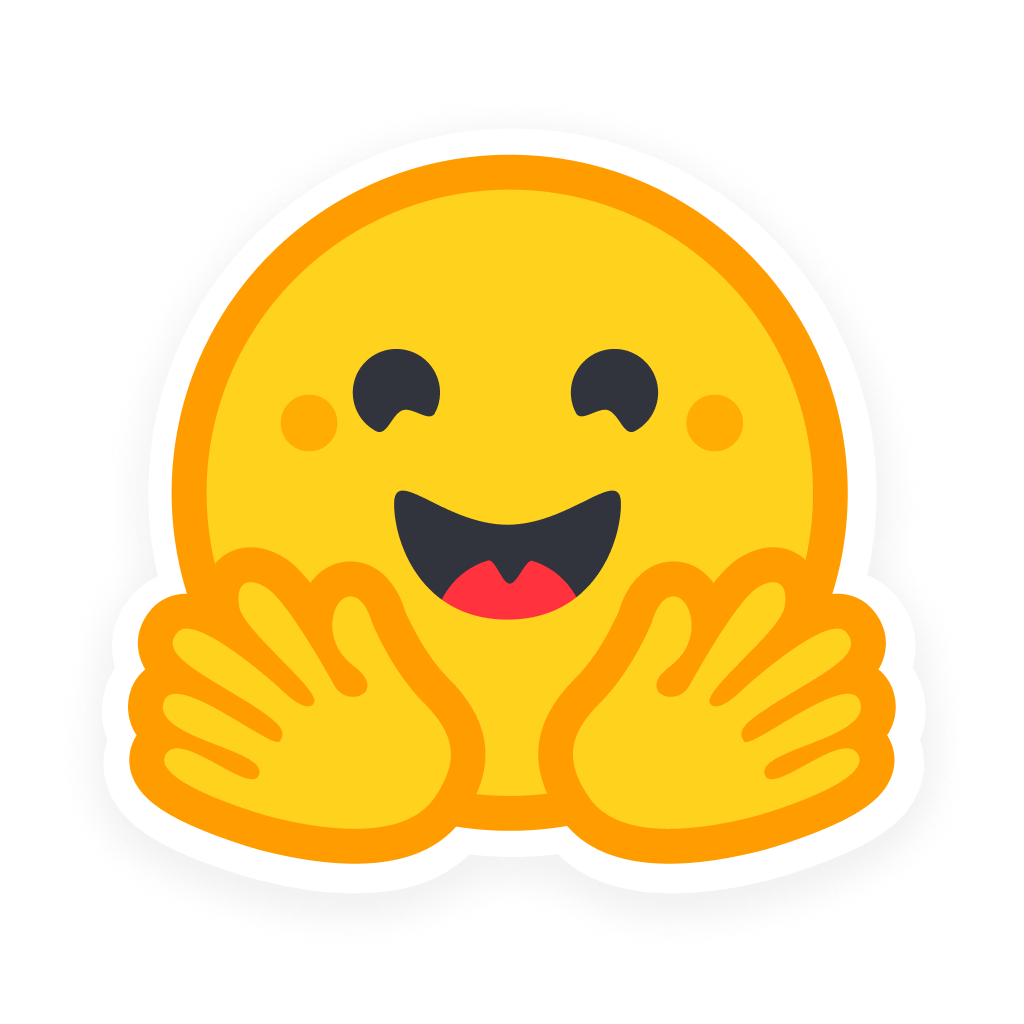}}\footnote{ \url{https://huggingface.co/mispeech/midashenglm-gen}}, and the demo page is available at \href{https://xingws.github.io/midashenglm-gen-demo/}{\includegraphics[height=1em,keepaspectratio]{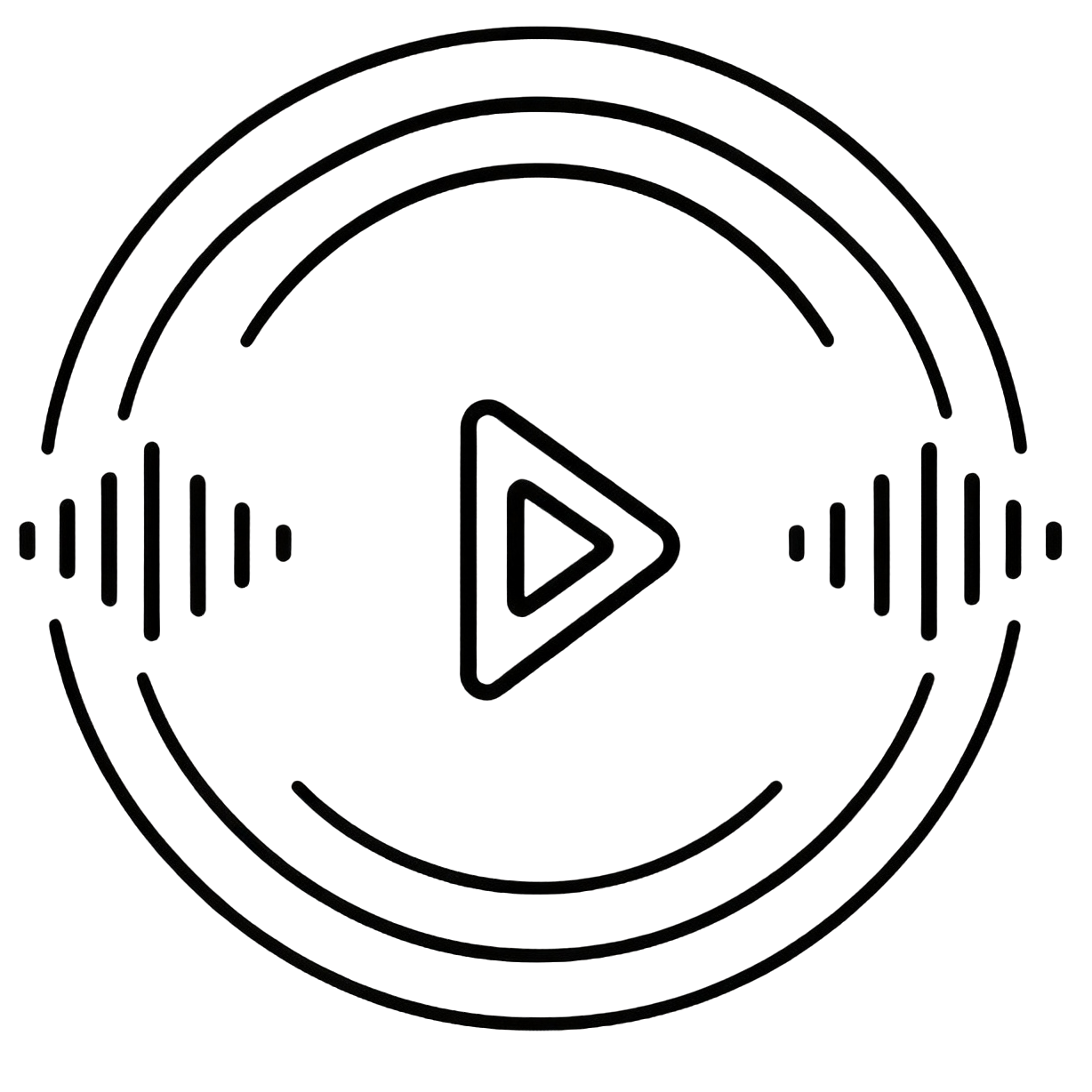}}\footnote{\url{https://xingws.github.io/midashenglm-gen-demo/}}.
\end{abstract}

\section{Introduction}

Generating coherent audio scenes that naturally blend intelligible speech, music, and environmental sounds is essential for applications such as film production, game design, and immersive media.
While single-type generation models have made remarkable progress in Text-to-Speech (TTS)~\cite{hu2026qwen3tts,seedtts}, Text-to-Music (TTM)~\cite{copet2023musicgen,agostinelli2023musiclm}, and Text-to-Audio (TTA)~\cite{liu2023audioldm,hung2024tangoflux}, they operate within isolated domains and cannot produce realistic mixed-audio scenes where multiple sound sources coexist and interact.
In contrast, task-unified models~\cite{chung2024uniaudio,xu2025uniflow} consolidate multiple generation tasks under a single architecture, but produce each audio type independently without modeling the temporal coordination, energy balance, and acoustic consistency required for coherent scenes where speech, music, and sound effects overlap.

Dasheng AudioGen~\cite{mei2026dashengaudiogen} made a first step toward this goal by introducing structured multi-view captions for fine-grained scene generation, decomposing complex acoustic scenes into textual descriptions and demonstrating that high-dimensional semantic-acoustic latent representations enable effective unified generation.
However, because Dasheng AudioGen's architecture relies on a frozen text encoder and a non-autoregressive diffusion transformer, the model is fundamentally limited to fixed-length, monolingual outputs with suboptimal speech intelligibility.
These limitations stem from fundamental architectural choices rather than the structured caption or latent space design, motivating an end-to-end generation backbone that jointly optimizes text understanding and audio synthesis.

% However, Dasheng AudioGen exhibits three key architectural limitations:
% \begin{enumerate}
%     \item By training only the diffusion transformer and keeping the text encoder entirely frozen, Dasheng AudioGen decouples text semantics from audio synthesis, which restricts data utilization and limits the model to monolingual English.
%     \item Consequently, the model exhibits suboptimal speech intelligibility, underperforming significantly on text-to-speech tasks compared to dedicated synthesis systems, a limitation potentially inherent to this diffusion-based approach.
%     \item The reliance on a non-autoregressive flow-matching diffusion transformer restricts generation to fixed-length audio outputs, thereby constraining its practical utility in varied deployment scenarios.
% \end{enumerate}

To address these limitations, we propose MiDashengLM-Gen, which retains the structured multi-view captions and the DashengTokenizer latent space from Dasheng AudioGen but replaces the generation backbone with a pre-trained large language model (LLM) combined with per-token conditional flow matching (Figure~\ref{fig:pipelines}).
The LLM jointly encodes text and audio, eliminating the separate frozen text encoder and thereby enabling multilingual generation and improved data utilization. This architecture brings speech intelligibility close to dedicated TTS systems while maintaining mixed-audio scene generation capability.
MiDashengLM-Gen combines LLM sequence modeling with per-token flow matching to enable autoregressive, variable-length, coherent mixed-audio scene generation within a unified framework.
\Cref{tab:generation_capabilities} compares the capabilities and architectural features of previous approaches.

\begin{table}[t]
\centering
\caption{Generation capabilities of representative audio generation models. Var.-Len. indicates whether the model supports variable-length generation.}
\label{tab:generation_capabilities}
\small
\begin{tabular}{l|cccccc}
\toprule
Model & Var.-Len. & SFX & Music & Speech & Scene & Multiling. \\
\midrule
TangoFlux~\cite{hung2024tangoflux} & \xmark & \cmark & \cmark & \xmark & \xmark & \naall \\
MusicGen~\cite{copet2023musicgen} & \cmark & \xmark & \cmark & \xmark & \xmark & \naall \\
Qwen3-TTS~\cite{hu2026qwen3tts} & \cmark & \xmark & \xmark & \cmark & \xmark & \cmark \\
UniFlow-Audio~\cite{xu2025uniflow} & \xmark & \cmark & \cmark & \cmark & \xmark & \xmark \\
Dasheng AudioGen~\cite{mei2026dashengaudiogen} & \xmark & \cmark & \cmark & \cmark & \cmark & \xmark \\
MiDashengLM-Gen & \cmark & \cmark & \cmark & \cmark & \cmark & \cmark \\
\bottomrule
\end{tabular}
\end{table}

% To systematically evaluate unified audio generation, we build a comprehensive evaluation pipeline covering both single-type and mixed-audio scene generation.
% In addition to standard benchmarks such as AudioCaps~\cite{kim2019audiocaps} and MusicCaps~\cite{agostinelli2023musiclm}, we evaluate on the Seed-TTS benchmark~\cite{seedtts}, a multilingual test set, and the mixed-audio MECAT benchmark.

Our contributions are:
\begin{enumerate}
    \item \textbf{Autoregressive generation with per-token flow matching.}
    We introduce MiDashengLM-Gen, a framework that integrates a pre-trained LLM backbone with per-token conditional flow matching for unified audio scene generation, supporting variable-length output via a learned stop head.

    \item \textbf{LLM-conditioned flow matching with high-dimensional audio latents.}
    We demonstrate the viability of utilizing high-dimensional semantic-acoustic latents for LLM-conditioned per-token flow matching. Furthermore, we identify a critical convergence prerequisite: the DiT decoder width must strictly exceed the audio latent dimensionality—a structural property that generalizes across LLM scales and latent dimensions.

    \item \textbf{Audio-text alignment pre-training for cross-modal generation.}
    We integrate an audio-text alignment stage that maps audio latents into the LLM's token space prior to generation training. Ablation studies confirm that this alignment phase is strictly required for viable cross-modal synthesis.

    \item \textbf{Strong speech intelligibility with competitive mixed-audio performance.}
    MiDashengLM-Gen achieves speech intelligibility approaching dedicated TTS systems, maintains competitive mixed-audio scene generation across multiple benchmarks, and supports multilingual generation.
\end{enumerate}

\section{Related Work}

\subsection{Single-Type Audio Generation}

\paragraph{Text-to-Audio (TTA).}
Generating sound effects from text descriptions has been primarily addressed through diffusion~\cite{ho2020denoising} and flow-matching~\cite{lipman2022flow} frameworks.
AudioLDM~\cite{liu2023audioldm} introduced CLAP-conditioned latent diffusion over compressed audio representations, and Make-An-Audio~\cite{huang2023makeanaudio} extended this paradigm with temporal structure modeling and prompt augmentation strategies.
TangoFlux~\cite{hung2024tangoflux} replaced iterative diffusion with single-step flow matching combined with preference-based training, significantly reducing inference cost while maintaining audio fidelity.
All these approaches target isolated environmental sounds and do not address speech or music generation.

\paragraph{Text-to-Music (TTM).}
Music generation presents unique challenges in long-term structural harmony, typically addressed via discrete-token or continuous-latent architectures.
MusicLM~\cite{agostinelli2023musiclm} pioneered a hierarchical approach that first generates semantic tokens and then refines them into acoustic tokens.
MusicGen~\cite{copet2023musicgen} simplified this pipeline by directly predicting EnCodec tokens with a single text-conditioned transformer.
On the continuous side, AudioLDM2~\cite{liu2023audioldm2} extends latent diffusion to a shared audio-language embedding space, and JEN-1~\cite{li2023jen} applies flow matching for controllable music creation.
Despite strong musical output, none of these systems can produce intelligible speech or model interactions between music and other concurrent sounds.

\paragraph{Text-to-Speech (TTS).}
The TTS field has shifted from pipeline-based synthesis such as Tacotron~\cite{shen2018tacotron} and FastSpeech~\cite{ren2019fastspeech} to fully end-to-end architectures like VITS~\cite{kim2021vits} that jointly model linguistic and acoustic features.
Discrete-token language models including VALL-E~\cite{wang2023valle} and SoundStorm~\cite{borsos2023soundstorm} then demonstrated that speech can be effectively cast as a sequence prediction task over audio codecs, opening the path for LLM-based TTS.
Building on this paradigm, large-scale systems such as Seed-TTS~\cite{seedtts}, Qwen3-TTS~\cite{hu2026qwen3tts}, and MiniMax-Speech~\cite{zhang2025minimaxspeech} achieve near-human quality with zero-shot voice cloning and multilingual support.
Most recently, flow-matching approaches including F5-TTS~\cite{chen2024f5tts} and CosyVoice 3~\cite{du2025cosyvoice3} offer efficient non-autoregressive synthesis with fine-grained prosody and emotion control.
However, all TTS systems produce speech in isolation, without accounting for background music, environmental acoustics, or other simultaneously occurring audio events.

\subsection{Mixed-Type Audio Generation}

\paragraph{Task-Specific Multi-Type Generation.}
Several architectures support generating different audio types but treat each as a separate task.
UniAudio~\cite{chung2024uniaudio} first unified multiple audio tasks by encoding all types as flat token sequences processed by a shared language model. AudioX~\cite{tian2025audiox} broadened the input space by conditioning a diffusion transformer on heterogeneous modalities beyond text.
UniFlow-Audio~\cite{xu2025uniflow} further adopted a common flow-matching backbone with modality-specific adapters for flexible generation.
Although these models consolidate multiple capabilities, they generate each audio type independently and cannot produce scenes where speech, music, and environmental sounds overlap and interact coherently.

\paragraph{Unified Audio Scene Generation.}
BagPiper~\cite{tian2026bagpiper} demonstrates mixed audio generation via an autoregressive framework, but relies on very long unstructured captions that make fine-grained control difficult.
Further, Dasheng AudioGen~\cite{mei2026dashengaudiogen} proposed structured multi-view captions to decompose scenes into distinct textual descriptions (global caption, transcript, music, sound events, and environment), and showed that generating in a high-dimensional semantic-acoustic latent space is more effective than using compressed VAE features.
However, its non-autoregressive architecture restricts output to a fixed duration and relies on a separate frozen text encoder, decoupling text semantics from audio synthesis.

In summary, existing models either generate audio types independently (UniAudio, UniFlow-Audio) or are constrained by frozen text encoders and fixed output durations (Dasheng AudioGen). To address these constraints, we integrate an LLM backbone that jointly optimizes text understanding and audio generation with per-token flow matching, allowing for flexible output lengths and coherent scene generation.

\section{Method}

\begin{figure}[t]
    \centering
    \begin{subfigure}[b]{0.48\linewidth}
        \centering
        \includegraphics[width=\linewidth]{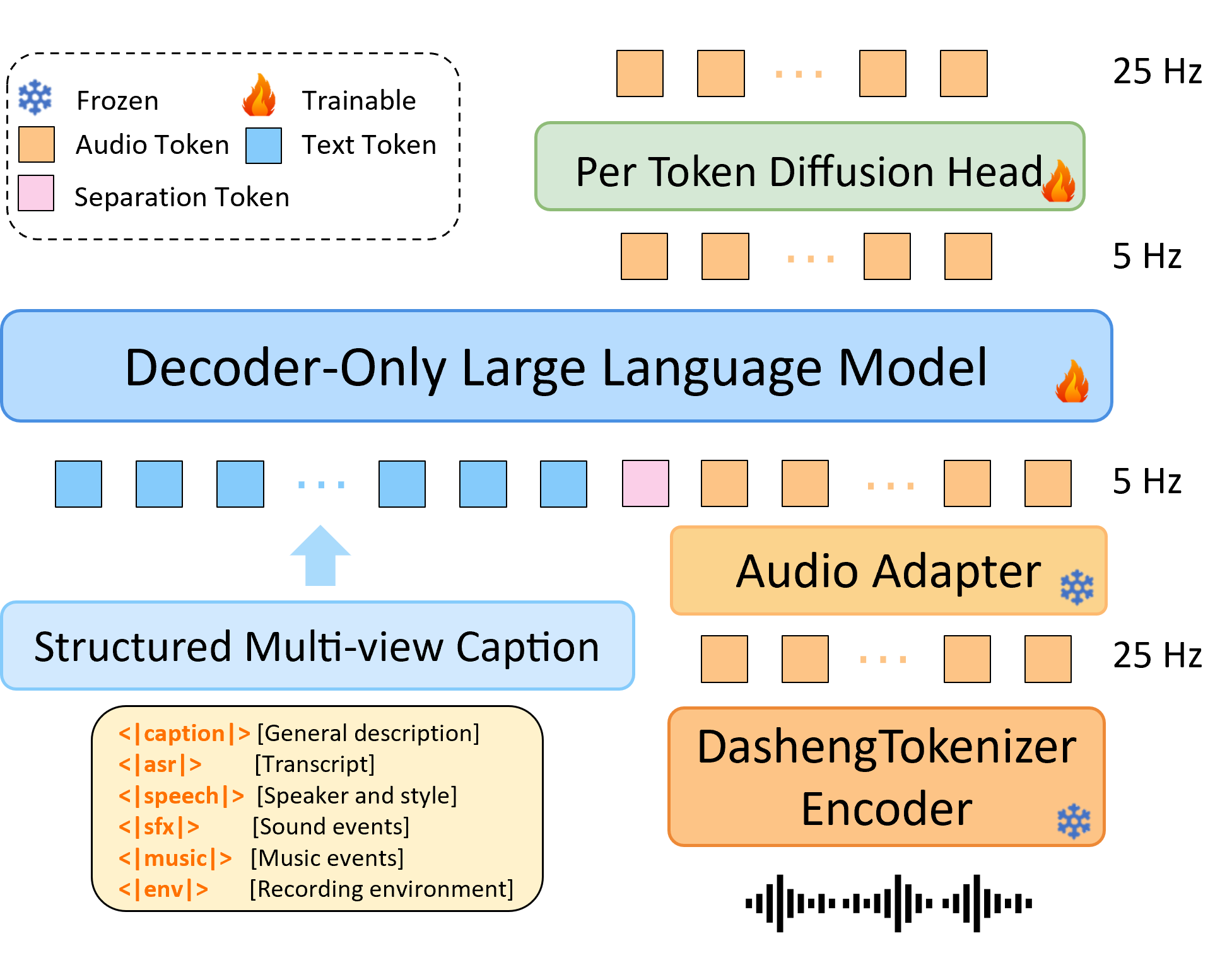}
        \caption{Training pipeline with flow matching loss.}
        \label{fig:training_pipeline}
    \end{subfigure}
    \hfill
    \begin{subfigure}[b]{0.48\linewidth}
        \centering
        \includegraphics[width=\linewidth]{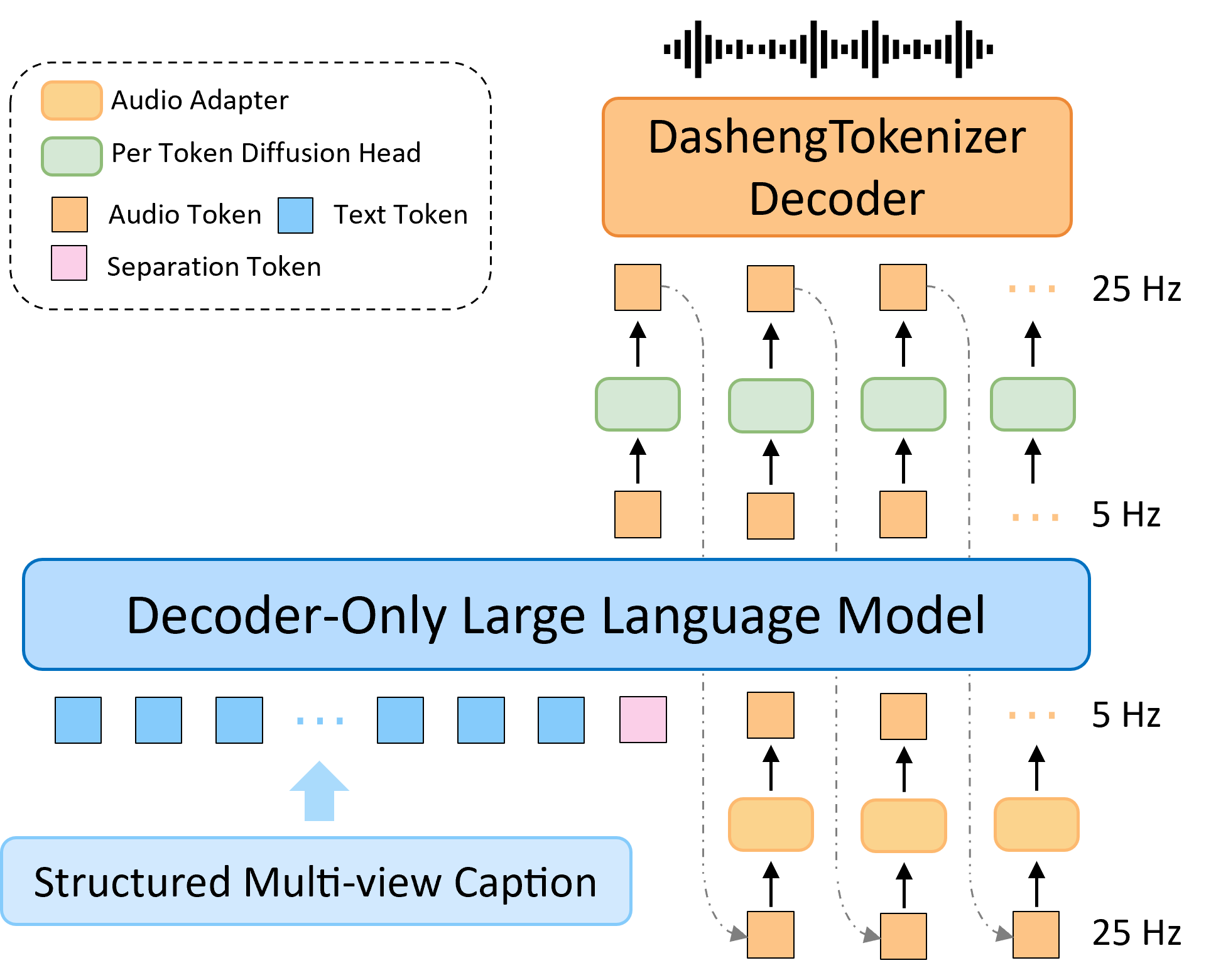}
        \caption{Inference pipeline with autoregressive generation.}
        \label{fig:inference_pipeline}
    \end{subfigure}
    \caption{Training and inference pipelines.}
    \label{fig:pipelines}
\end{figure}

Our goal is to generate coherent audio scenes from text descriptions, where the scene may simultaneously contain speech, music, sound effects, and environmental acoustics. 
We formulate this as autoregressive conditional generation: the model produces audio tokens sequentially, with each token conditioned on the text description and all previously generated audio latents. 
Our framework combines autoregressive LLM sequence modeling with per-token continuous flow matching. 
A DiT-based flow matching decoder generates continuous audio tokens, conditioned on the LLM hidden state at each autoregressive step. 
We adopt the encoder of a custom MiDashengLM-0.6B model\footnote{\url{https://huggingface.co/mispeech/midashenglm-0.6b-fp32}} as our unified semantic-acoustic audio tokenizer, following the DashengTokenizer~\cite{dinkel2026dashengtokenizer} architecture. 
Structured multi-view captions provide fine-grained supervision across six textual views.
\Cref{fig:training_pipeline,fig:inference_pipeline} shows the overall model architecture.

\subsection{Autoregressive LLM Backbone}
\label{ssec:llm_backbone}

We employ a pre-trained causal language model (Qwen3~\cite{yang2025qwen3}) as the sequential backbone for audio scene generation. 
Unlike prior unified models that require separate text encoders (e.g., T5) and cross-attention mechanisms, our approach uses the LLM's pre-trained representations to jointly encode text and audio within a single unified model. 
This eliminates architectural complexity while leveraging the LLM's strong language understanding for text-to-audio alignment.

\paragraph{Structured Multi-View Audio Scene Captioning}
To provide fine-grained supervision for complex audio scenes, we decompose each scene into six complementary textual views: global scene description, speech transcript, speaker identity and style, sound effects, music content, and recording environment. 
Each view is associated with a dedicated special token (e.g., \texttt{<|caption|>}, \texttt{<|asr|>}, \texttt{<|speech|>}, \texttt{<|sfx|>}, \texttt{<|music|>}, \texttt{<|env|>}). 
The general description must be included, while remaining fields are used only when applicable, otherwise filled with \texttt{<|unknown|>}. 
For example, pure speech samples fill \texttt{<|music|>} and \texttt{<|sfx|>} with \texttt{<|unknown|>}.
This structured conditioning exposes different semantic views to the LLM via explicit special tokens, reducing semantic entanglement among control factors.

\begin{figure}[t]
    \centering
    \includegraphics[width=0.48\linewidth]{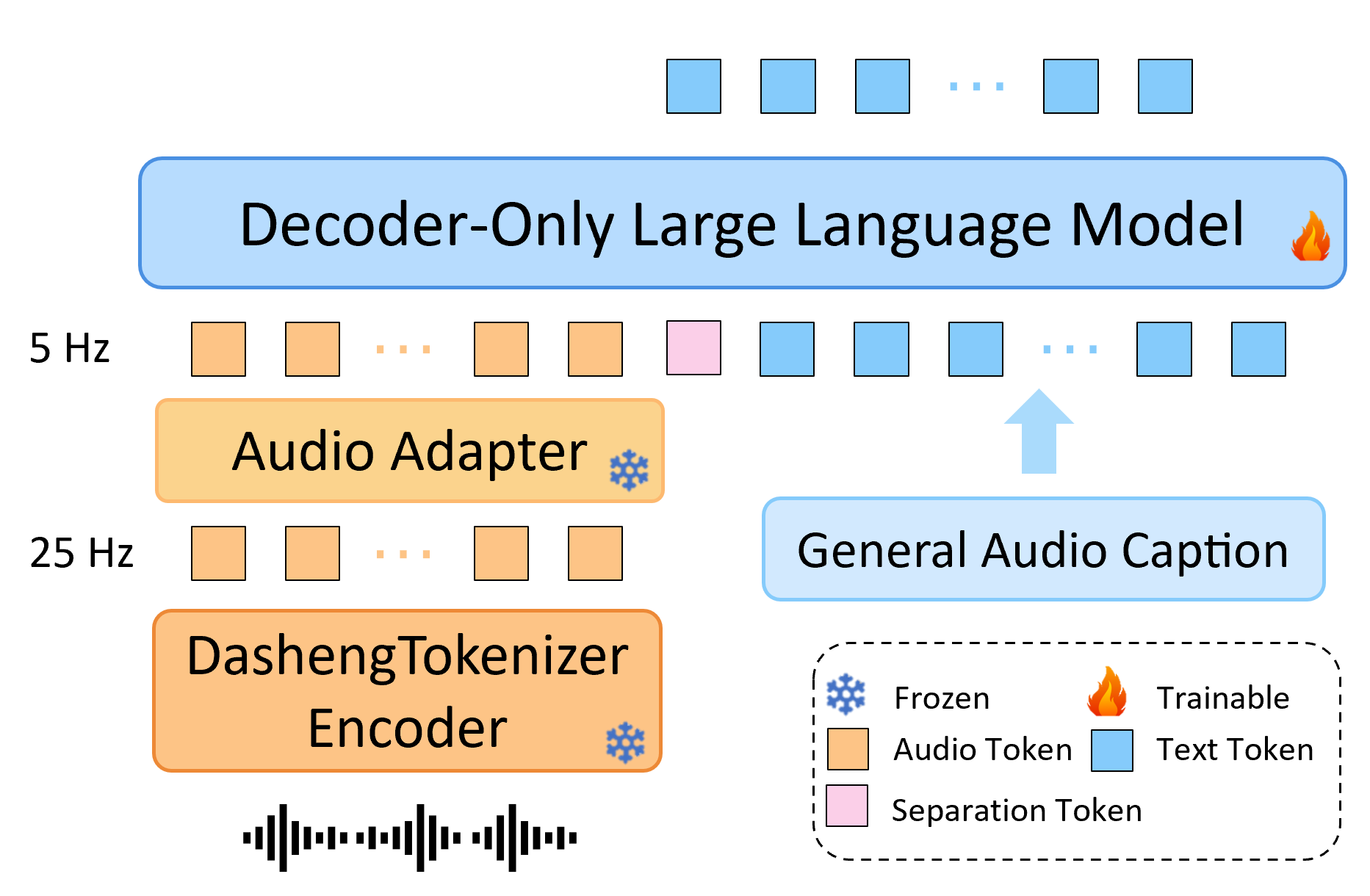}
    \caption{Audio-text alignment in the DashengTokenizer latent space via general caption supervision.}
    \label{fig:modal_alignment}
\end{figure}

\paragraph{Audio-Text Alignment}\label{ssec:audio_text_alignment} Before training the generation model, we first align audio token latents with the LLM's text embedding space. 
As we show, without explicit alignment, the LLM cannot effectively process the complex audio scenes necessary for generation. 
Figure~\ref{fig:modal_alignment} illustrates this alignment process.
Following MiDashengLM~\cite{dinkel2025midashenglm}, we perform audio-text alignment via general audio captions. 
We use the encoder of a custom MiDashengLM-0.6B model as our unified semantic-acoustic latent space, which produces continuous representations at 25\,Hz. 
Unlike low-dimensional acoustic VAE latents~\cite{liu2023audioldm2,xu2025uniflow}, DashengTokenizer preserves both semantic information and acoustic detail, shortening the cross-modal mapping from text to audio.

During alignment training, an audio adapter maps the DashengTokenizer's latents into the LLM's token space. To boost efficiency, we downsample the audio sequence to 5\,Hz by grouping $k=5$ consecutive encoder frames and projecting them via an MLP:
\begin{equation}
\mathbf{a}_i = \mathrm{MLP}\left(\mathrm{concat}(\mathbf{z}_{(i-1)k+1}, \ldots, \mathbf{z}_{ik})\right) \in \mathbb{R}^{d_{\text{llm}}},
\end{equation}
where $\mathbf{z}$ is the audio latent from the encoder of DashengTokenizer and $\mathbf{a}_i \in \mathbb{R}^{d_{\text{llm}}}$ is the projected audio token latent at 5\,Hz (200\,ms per token) for LLM input.

\paragraph{Autoregressive LLM Input Formulation}
Given a structured multi-view caption, we construct an input sequence by concatenating the caption text and a sequence of audio token latents:
\begin{equation}
\mathbf{h} = \mathrm{LLM}\left([\mathbf{e}_{\text{caption}}, \texttt{<|audio\_bos|>}, \mathbf{a}_1, \mathbf{a}_2, \ldots, \mathbf{a}_N]\right),
\end{equation}
where $\mathbf{e}_{\text{caption}}$ is the text embedding of the structured caption, $\texttt{<|audio\_bos|>}$ marks the start of audio generation, $\mathbf{a}_i$ represents the audio latents.
The hidden state $\mathbf{h}_i \in \mathbb{R}^{d_{\text{llm}}}$ at each audio position $i$ encodes the full context (text semantics, prior audio history, and positional information) and serves as the condition for generating the $i$-th audio token via per-token flow matching (\Cref{ssec:per_token_flow}).
During training, ground-truth audio latents are fed at each position.

\subsection{Per-Token Conditional Flow Matching}
\label{ssec:per_token_flow}

Unlike full-sequence flow matching that generates the entire audio latent sequence in one pass we perform conditional flow matching independently for each audio token, conditioned on the LLM hidden state $\mathbf{h}_i$. 
This per-token design preserves the autoregressive structure's adaptive length control while retaining the expressiveness of continuous flow-based generation. 
Compared to discrete-token autoregressive approaches (e.g., VALL-E~\cite{wang2023valle}), operating over continuous high-dimensional latents avoids quantization artifacts and supports richer acoustic modeling.

\paragraph{Token-Level Generation.}
For each audio token position $i$, the flow-matching DiT generates a group of $k=5$ encoder frames $\hat{\mathbf{z}}_{(i-1)k+1:ik} \in \mathbb{R}^{k \times d_a}$, by processing the concatenation of latent history frames $\hat{\mathbf{z}}_{(i-2)k+1:(i-1)k}$, and noised target frames, $\mathbf{z}_0^{(i)} \sim \mathcal{N}(0, I)$, conditioned on the timestep embedding and LLM hidden state $\mathbf{h}_i \in \mathbb{R}^{d_{\text{llm}}}$.
Each DiT block applies self-attention with rotary position embeddings (RoPE) followed by a feed-forward network. 

\paragraph{Training Objective.}
Let $\mathbf{z}_1^{(i)} \in \mathbb{R}^{k \times d_a}$ denote the ground-truth frames for token $i$, produced by the DashengTokenizer encoder at 25\,Hz without audio adapter projection. For a random timestep $t \sim \mathcal{U}(0,1)$, we construct the interpolated sample $\mathbf{z}_t^{(i)} = (1-t)\mathbf{z}_0^{(i)} + t\,\mathbf{z}_1^{(i)}$, where $\mathbf{z}_0^{(i)} \sim \mathcal{N}(0, I)$ is Gaussian noise. 
The DiT learns to predict the conditional vector field:
\begin{equation}
\mathcal{L}_{\mathrm{FM}}
=
\mathbb{E}_{i, \mathbf{z}_0^{(i)}, \mathbf{z}_1^{(i)}, t}
\left[
\left\|
v_\theta(\mathbf{z}_t^{(i)}, t, \mathbf{h}_i, \mathbf{z}_{\text{hist}}^{(i)}) - (\mathbf{z}_1^{(i)}-\mathbf{z}_0^{(i)})
\right\|_2^2
\right],
\end{equation}
where $\mathbf{h}_i$ is the LLM hidden state at position $i$, $\mathbf{z}_{\text{hist}}^{(i)}$ contains the latent history frames, and the loss is masked to only include valid (non-padded) audio positions.

\paragraph{Inference.}
At inference, we solve the ODE from noise to signal using a 10-step Euler solver. To improve conditioning strength, we apply classifier-free guidance (CFG) by interpolating between unconditional and conditional predictions:
\begin{equation}
\hat{v} = v_\theta(\mathbf{z}_t, t, \varnothing, \mathbf{z}_{\text{hist}}) + w \cdot \left(v_\theta(\mathbf{z}_t, t, \mathbf{h}_i, \mathbf{z}_{\text{hist}}) - v_\theta(\mathbf{z}_t, t, \varnothing, \mathbf{z}_{\text{hist}})\right),
\end{equation}
where $\varnothing$ is a learnable null embedding and $w$ is the guidance scale. During training, we randomly replace the LLM condition with the null embedding to enable CFG at inference.

Generation length is determined by a learned stop head that predicts a binary continue/stop distribution over LLM hidden states, trained with balanced cross-entropy (hard negatives mined from lowest-confidence non-stop tokens) and combined with $\mathcal{L}_{\mathrm{FM}}$ via weight $\lambda$. At inference, generation halts when the stop probability exceeds threshold $\tau$ after a minimum of $i_{\min}$ steps. Audio is decoded from the generated latent sequence by the DashengTokenizer decoder: a 5$\times$ convolutional upsampler followed by a Vocos~\cite{siuzdak2024vocos} vocoder at 16\,kHz.

\subsection{Implementation Details}
\label{ssec:appendix_implementation_details}

\paragraph{Model Configuration.}
We use Qwen3-1.7B~\cite{yang2025qwen3} as the LLM backbone with full fine-tuning. 
We extract the encoder from a custom MiDashengLM-0.6B model, producing 768-dimensional latents at 25\,Hz.
The audio projector downsamples to 5\,Hz tokens projected to the LLM's 2048-dimensional space. 
The DashengTokenizer encoder, audio projector, upsampler, and decoder are frozen during the main training phase and the LLM, DiT, and stop head are trained. 
The DiT in flow matching has 16 layers, hidden dimension 2048, 8 attention heads, and MLP ratio of 4.0.

\paragraph{Training and Inference.}
We train with AdamW (8-bit)~\cite{dettmers2022llmint8} with the following hyperparameters: batch size 16, learning rate $5 \times 10^{-5}$ with cosine decay to $1 \times 10^{-6}$, and gradient clipping at 1.0. 
Training runs for 800 epochs of 2000 iterations each. 
Audio inputs range from 1 to 20 seconds at 16\,kHz. 
The stop loss weight is $\lambda = 0.01$, stop threshold $\tau = 0.5$, and minimum steps $i_{\min} = 5$ (1 second) at inference. 
CFG scale is 2.0 with 10 ODE steps.

\section{Experiments}

\subsection{Experimental Setup}

\noindent\textbf{Training Data.}
Our training data comprises two sources.
The first is a private superset of ACAVCaps~\cite{acavcaps} with 77k hours of audio covering speech, music, and sound effects, derived from ACAV100M~\cite{lee2021acav100m}.
ACAVCaps uses a multi-expert annotation pipeline that analyzes each audio clip from six domain-specific perspectives; we convert these annotations into our structured multi-view caption format.
Audio samples are categorized into single-type (S00 = speech-only, 0M0 = music-only, 00A = sound effects) and mixed categories (0MA, S0A, SM0, SMA), where S, M, and A denote the presence of speech, music, and sound effects, respectively, and 0 denotes absence.
The second source consists of dedicated TTS corpora to strengthen speech synthesis: Emilia~\cite{he2024emilia} (Chinese and English subsets), LibriTTS~\cite{LibriTTS}, LJSpeech~\cite{ljspeech17}, AISHELL-3~\cite{shi2020aishell3}, and WenetSpeech4TTS~\cite{ma2024wenetspeech4tts}.
For TTS data, we prepend a general caption (``natural human speech'') to align with the multi-view caption format.
Table~\ref{tab:training_data} details the language distribution of training data.
\begin{table}[t]
  \centering
  \caption{Language distribution of training data. The ACAVCaps column reports the speech-only (S00) subset; TTS data is predominantly English and Chinese.}
  \label{tab:training_data}
  \small
  \begin{tabular}{lccccc}
  \toprule
  & \multicolumn{2}{c}{ACAVCaps (S00)} & \multicolumn{2}{c}{TTS} & Overall \\
  \cmidrule(lr){2-3} \cmidrule(lr){4-5} \cmidrule(lr){6-6}
  Language & Duration (h) & Proportion (\%) & Duration (h) & Proportion (\%) & Proportion (\%) \\
  \midrule
  English & 15{,}368 & 58.86 & 11{,}074 & 51.90 & 53.82 \\
  Chinese & 70 & 0.27 & 11{,}949 & 48.10 & 24.46 \\
  Spanish & 2{,}741 & 10.50 & -- & -- & 5.58 \\
  Portuguese & 1{,}916 & 7.34 & -- & -- & 3.90 \\
  Russian & 1{,}217 & 4.66 & -- & -- & 2.48 \\
  French & 934 & 3.58 & -- & -- & 1.90 \\
  Japanese & 875 & 3.35 & -- & -- & 1.78 \\
  Korean & 848 & 3.25 & -- & -- & 1.73 \\
  German & 842 & 3.23 & -- & -- & 1.71 \\
  Other & 1{,}299 & 4.98 & -- & -- & 2.64 \\
  \bottomrule
  \end{tabular}
\end{table}

\noindent\textbf{Evaluation Benchmarks.}
For single-type audio and music generation, we report on AudioCaps~\cite{kim2019audiocaps} and MusicCaps~\cite{agostinelli2023musiclm}.
For speech generation, we evaluate on the Seed-TTS benchmark~\cite{seedtts} for English and Chinese intelligibility, and adopt the multilingual test set from MiniMax-Speech~\cite{zhang2025minimaxspeech} covering 9 languages.
We report CERs (Character Error Rate) for Chinese, Japanese, and Korean, and WERs (Word Error Rate) for all other languages. 
% TODO ?????? 
Chinese speech is transcribed using Paraformer-zh~\cite{gao2023funasr}, while all other languages are transcribed using Whisper-large-v3~\cite{radford2023whisper}.
For emotion expressiveness, we use the CV3-Eval Emotional Voice Cloning subset~\cite{du2025cosyvoice3}, which retains samples labeled as happy, sad, or angry (100 per language), split into text-related and text-unrelated settings depending on whether the target text is semantically consistent with the target emotion. 
Emotion classification accuracy is measured using the emotion2vec model~\cite{ma2024emotion2vec}.
Our primary benchmark for mixed audio scene generation is MECAT~\cite{nuu2025mecat}, a held-out test set of ACAVCaps that contains both single-type and mixed-type audio samples with rich multi-view annotations. 
Since MECAT is a multilingual benchmark whose speech categories contain utterances in multiple languages, we report objective results on the English subset\footnote{The MECAT English subset list is available at \url{https://github.com/xiaomi-research/dasheng-audiogen/tree/main/evaluation}.} of MECAT for speech-related categories to ensure fair comparison with baseline methods.
%TODO write out all abbrev
Across all benchmarks, we report audio distribution metrics: Fréchet Audio Distance (FAD)~\cite{kilgour2019fad}, Fréchet Distance (FD), and Kullback–Leibler divergence (KL)~\cite{liu2023audioldm}, text similarity metrics: Contrastive Language-Audio Pretraining (CLAP)~\cite{elizalde2022clap}, and speech-related metrics: WER/CER. In all tables, \textbf{bold} and \underline{underlined} denote the best and second-best results, respectively.

\begin{table}[t]
  \centering
  \caption{Results on AudioCaps and MusicCaps.}
  \label{tab:standard_benchmarks_updated}
  \small
  \setlength{\tabcolsep}{4pt}
  \begin{tabular}{lcccccccc}
  \toprule
  & \multicolumn{4}{c}{AudioCaps} & \multicolumn{4}{c}{MusicCaps} \\
  \cmidrule(lr){2-5} \cmidrule(lr){6-9}
  Method & $\mathrm{FAD}_{\mathrm{VGG}}$ $\downarrow$ & $\mathrm{FD}_{\mathrm{PANNS}}$ $\downarrow$ & KL $\downarrow$ & CLAP $\uparrow$ & $\mathrm{FAD}_{\mathrm{VGG}}$ $\downarrow$ & $\mathrm{FD}_{\mathrm{PANNS}}$ $\downarrow$ & KL $\downarrow$ & CLAP $\uparrow$ \\
  \midrule
  TangoFlux & \best{2.26} & \second{19.13} & \best{1.19} & \best{0.58} & \naall & \naall & \naall & \naall \\
  MusicGen & \naall & \naall & \naall & \naall & 3.80 & \naall & \second{1.31} & 0.28 \\
  UniFlow-Audio & 5.74 & \best{17.18} & \second{1.43} & \second{0.48} & 4.05 & 27.12 & 1.87 & 0.24 \\
  Dasheng AudioGen & \second{3.19} & 26.06 & 1.86 & 0.44 & \best{1.37} & \second{18.45} & 1.37 & \best{0.33} \\
  MiDashengLM-Gen & 5.01 & 26.13 & 1.91 & 0.36 & \second{2.28} & \best{14.58} & \best{1.24} & \best{0.33} \\
  \bottomrule
  \end{tabular}
  \end{table}

\begin{table}[t]
  \centering
  \caption{Speech intelligibility on the Seed-TTS benchmark.}
  \label{tab:tts_results}
  \small
  \begin{tabular}{lcc}
  \toprule
  Method & EN WER(\%) $\downarrow$ & ZH CER(\%) $\downarrow$ \\
  \midrule
  Seed-TTS & \second{2.25} & \second{1.12} \\
  Qwen3-TTS & \best{1.24} & \best{0.77} \\
  Dasheng AudioGen & 12.15 & $>$100 \\
  MiDashengLM-Gen & 2.79 & 3.87 \\
  \bottomrule
  \end{tabular}
  \end{table}

\begin{table}[t]
  \centering
  \caption{Multilingual TTS evaluation (WER/CER\% $\downarrow$).}
  \label{tab:multilingual_tts}
  \small
  \begin{tabular}{lcccc}
  \toprule
  Language & MiniMax & ElevenLabs & Dasheng AudioGen & MiDashengLM-Gen \\
  \midrule
  English & \best{2.16} & \second{2.34} & 8.60 & 2.42 \\
  Chinese & \best{2.25} & 16.03 & 99.62 & \second{3.51} \\
  Spanish & \best{1.03} & \second{1.08} & 9.31 & 3.11 \\
  Portuguese & \second{1.88} & \best{1.33} & 8.98 & 8.04 \\
  Russian & \second{4.28} & \best{3.88} & 36.34 & 13.19 \\
  French & \best{4.10} & \second{5.22} & 17.96 & 12.53 \\
  Japanese & \best{3.52} & \second{10.65} & 109.21 & 16.45 \\
  Korean & \best{1.75} & \second{1.87} & 100.25 & 5.49 \\
  German & \second{1.91} & \best{0.57} & 24.15 & 4.38 \\
  \bottomrule
  \end{tabular}
\end{table}

\begin{table}[t]
  \centering
  \caption{Emotion expressiveness on CV3-Eval (accuracy $\uparrow$).}
  \label{tab:emotion_eval}
  \small
  \begin{tabular}{lcccccc}
  \toprule
  & \multicolumn{3}{c}{Text-Related} & \multicolumn{3}{c}{Text-Unrelated} \\
  \cmidrule(lr){2-4} \cmidrule(lr){5-7}
  Method & Happy & Sad & Angry & Happy & Sad & Angry \\
  \midrule
  F5-TTS & 0.92 & 0.52 & 0.72 & \second{0.80} & 0.28 & \second{0.64} \\
  CosyVoice3-1.5B & 0.86 & 0.64 & 0.72 & 0.64 & 0.44 & 0.48 \\
  CosyVoice3$_{\text{DiffRO-EMO}}$ & \best{0.98} & \second{0.68} & \second{0.84} & \best{0.98} & \second{0.50} & \best{0.68} \\
  MiDashengLM-Gen & \best{0.98} & \best{0.96} & \best{0.92} & 0.78 & \best{0.52} & 0.62 \\
  \bottomrule
  \end{tabular}
\end{table}

\subsection{Sound Effect and Music Generation}
\label{ssec:standard_benchmarks}

To validate our approach's capabilities on single-type audio generation, we report results on AudioCaps and MusicCaps in~\Cref{tab:standard_benchmarks_updated}.
We compare against TangoFlux~\cite{hung2024tangoflux}, MusicGen~\cite{copet2023musicgen}, UniFlow-Audio~\cite{xu2025uniflow}, and Dasheng AudioGen~\cite{mei2026dashengaudiogen}.
On AudioCaps, our model trails the task-optimized TangoFlux (FAD 2.26 vs.\ 5.01).
This gap is expected: DashengLM-Gen is trained on mixed-audio scenes and TTS corpora, not on pure sound-effect data, and its architecture prioritizes cross-modal alignment and multi-source coordination over single-source acoustic fidelity.
TangoFlux, by contrast, is a dedicated text-to-audio model trained specifically on sound-effect corpora and uses a non-autoregressive flow-matching architecture that can optimize acoustic quality globally in a single pass.

On MusicCaps, our approach achieves notably better FD (14.58 vs.\ 18.45) and KL (1.24 vs.\ 1.37) compared to Dasheng AudioGen, demonstrating that the LLM backbone provides stronger cross-modal alignment for music generation through causal attention over longer contexts.

\subsection{Speech Generation}

\paragraph{Speech Generation on Seed-TTS.}
Table~\ref{tab:tts_results} reports WER on the Seed-TTS benchmark~\cite{seedtts}, which evaluates speech intelligibility on English and Chinese test sets. Our approach achieves 2.79\% on Seed-EN and 3.87\% on Seed-ZH, substantially improving over Dasheng AudioGen (12.15\% and $>$100\%, respectively). This demonstrates that the LLM backbone leverages its pre-trained language understanding to produce intelligible speech. A 2.2$\times$ gap remains against specialized TTS systems such as Qwen3-TTS (1.24\%/0.77\%) and Seed-TTS (2.25\%/1.12\%), primarily because our model is trained on mixed-audio scenes rather than optimized specifically for speech intelligibility.

\paragraph{Multilingual Speech Generation.}
To evaluate multilingual capability, Table~\ref{tab:multilingual_tts} reports WER/CER across 9 languages. Our model achieves competitive WER in high-resource languages (2.42\% English, 3.51\% Chinese), where training data is abundant. For under-represented languages such as Japanese (16.45\%) and Russian (13.19\%), the higher WER is consistent with the training data distribution shown in Table~\ref{tab:training_data}, where these languages have limited representation in the TTS corpus. Compared to Dasheng AudioGen, which suffers from extremely high error rates on most languages (e.g., 99.62\% Chinese, 109.21\% Japanese), our LLM-based approach demonstrates dramatically better multilingual speech intelligibility.

\paragraph{Emotion Expressiveness.}
\label{ssec:emotion}

To evaluate emotion control capability, we conduct experiments on CV3-Eval, which measures emotion classification accuracy under two settings: \emph{text-related} (emotion cue present in the text prompt) and \emph{text-unrelated} (emotion conveyed purely through prosody without textual cues). Table~\ref{tab:emotion_eval} reports results for three emotions (happy, sad, angry).

Our approach achieves the highest text-related emotion accuracy across all three emotions (0.98/0.96/0.92), demonstrating that the LLM backbone effectively conditions on emotional cues in the text prompt to generate appropriately expressive speech. In text-unrelated scenarios, where emotion must be conveyed solely through prosody, CosyVoice3$_{\text{DiffRO-EMO}}$ achieves the strongest performance (0.98/0.50/0.68), suggesting that its specialized emotion control mechanism better preserves prosodic emotion without textual cues. Our model achieves the best sad accuracy (0.52) in this setting, indicating competitive emotion expressiveness even without explicit textual guidance.

\subsection{MECAT Benchmark}
\label{ssec:mecat}

\noindent To comprehensively evaluate model performance in complex mixed-audio scenes, we conduct experiments on the MECAT benchmark. We compare against TangoFlux~\cite{hung2024tangoflux}, MusicGen~\cite{copet2023musicgen}, Qwen3-TTS~\cite{hu2026qwen3tts}, and Dasheng AudioGen~\cite{mei2026dashengaudiogen}.

\noindent{\textbf{Single-Type Categories.}} 
Table~\ref{tab:mecat_single} compares models on MECAT single-type categories. On the sound effects category (00A), Dasheng AudioGen achieves better FAD (4.25 vs.\ 6.05), while our approach achieves better FD (18.74 vs.\ 19.37) and KL (1.32 vs.\ 1.36). On the music category (0M0), our approach demonstrates a significant advantage in FD (6.82 vs.\ 14.90), reducing the distance by more than 2$\times$, while maintaining competitive FAD (1.97 vs.\ 1.66) and KL (0.57 vs.\ 0.61).

On the speech category (S00), our approach achieves substantially better FAD (0.87 vs.\ 1.76) and CLAP (0.35 vs.\ 0.34), indicating stronger alignment between generated speech and its acoustic context. The LLM backbone enables better modeling of speech together with environmental details, whereas Dasheng AudioGen generates speech more detached from the scene.

\begin{table}[t]
  \centering
  \caption{Results on MECAT single-type categories.}
  \label{tab:mecat_single}
  \small
  \begin{tabular}{llcccc}
  \toprule
  Category & Method & $\mathrm{FAD}_{\mathrm{VGG}}$ $\downarrow$ & $\mathrm{FD}_{\mathrm{PANNS}}$ $\downarrow$ & KL $\downarrow$ & CLAP $\uparrow$ \\
  \midrule
   \multirow{5}{*}{00A} & TangoFlux & 15.68 & 48.81 & 2.30 & 0.30 \\
   & MusicGen & 43.68 & 87.41 & 5.99 & -0.03 \\
   & Qwen3-TTS & 51.42 & 153.55 & 5.90 & -0.04 \\
   & Dasheng AudioGen & \best{4.25} & \second{19.37} & \second{1.36} & \best{0.37} \\
   & MiDashengLM-Gen & \second{6.05} & \best{18.74} & \best{1.32} & \second{0.35} \\
  \midrule
  \multirow{5}{*}{0M0} & TangoFlux & 5.04 & 24.97 & 0.82 & \best{0.32} \\
   & MusicGen & 4.53 & 19.75 & 0.88 & 0.26 \\
   & Qwen3-TTS & 21.68 & 137.10 & 4.99 & -0.03 \\
   & Dasheng AudioGen & \best{1.66} & \second{14.90} & \second{0.61} & 0.27 \\
   & MiDashengLM-Gen & \second{1.97} & \best{6.82} & \best{0.57} & \second{0.28} \\
  \midrule
  \multirow{5}{*}{S00} & TangoFlux & 10.54 & 53.15 & 1.63 & 0.29 \\
   & MusicGen & 26.74 & 89.71 & 5.38 & 0.04 \\
   & Qwen3-TTS & 8.75 & 27.55 & 0.84 & 0.31 \\
   & Dasheng AudioGen & \second{1.76} & \best{3.93} & \best{0.40} & \second{0.34} \\
   & MiDashengLM-Gen & \best{0.87} & \second{4.15} & \second{0.44} & \best{0.35} \\
  \bottomrule
  \end{tabular}
  \end{table}

\noindent{\textbf{Mixed Categories.}} 
Table~\ref{tab:mecat_mixed_updated} reports results on the mixed-audio categories. Across all speech-containing mixed categories (S0A, SM0, SMA), our approach consistently achieves lower FAD than Dasheng AudioGen: 1.54 vs.\ 1.75 on S0A, 0.98 vs.\ 1.70 on SM0, and 1.88 vs.\ 2.17 on SMA. However, Dasheng AudioGen retains advantages on feature-level distribution metrics (FD and KL) across these categories, suggesting its non-autoregressive architecture better captures fine-grained acoustic details. The overall tradeoff indicates that the LLM backbone improves multi-source coordination and spectral quality, while the non-autoregressive DiT has an edge in distributional fidelity for isolated aspects of the audio.

On the non-speech mixed category (0MA), Dasheng AudioGen achieves better FAD (3.86 vs.\ 4.23), while our approach achieves better FD (28.32 vs.\ 31.13) and KL (1.25 vs.\ 1.36).

\begin{table}[t]
  \centering
  \caption{Results on MECAT mixed-audio categories.}
  \label{tab:mecat_mixed_updated}
  \small
  \begin{tabular}{llcccc}
  \toprule
  Category & Method & $\mathrm{FAD}_{\mathrm{VGG}}$ $\downarrow$ & $\mathrm{FD}_{\mathrm{PANNS}}$ $\downarrow$ & KL $\downarrow$ & CLAP $\uparrow$ \\
  \midrule
   \multirow{5}{*}{0MA} & TangoFlux & 9.71 & 58.44 & 2.45 & \best{0.32} \\
   & MusicGen & 13.06 & 60.98 & 2.86 & 0.19 \\
   & Qwen3-TTS & 23.17 & 149.24 & 5.28 & -0.03 \\
   & Dasheng AudioGen & \best{3.86} & \second{31.13} & \second{1.36} & 0.31 \\
   & MiDashengLM-Gen & \second{4.23} & \best{28.32} & \best{1.25} & \best{0.32} \\
  \midrule
  \multirow{5}{*}{S0A} & TangoFlux & 10.92 & 43.94 & 2.23 & \best{0.36} \\
   & MusicGen & 30.38 & 89.09 & 5.60 & 0.00 \\
   & Qwen3-TTS & 19.62 & 62.84 & 2.42 & 0.15 \\
   & Dasheng AudioGen & \second{1.75} & \best{8.56} & \best{0.69} & \best{0.36} \\
   & MiDashengLM-Gen & \best{1.54} & \second{8.84} & \second{0.72} & \best{0.36} \\
  \midrule
  \multirow{5}{*}{SM0} & TangoFlux & 11.57 & 40.77 & 1.11 & \best{0.34} \\
   & MusicGen & 19.83 & 57.28 & 2.38 & 0.14 \\
   & Qwen3-TTS & 10.20 & 63.15 & 2.16 & 0.19 \\
   & Dasheng AudioGen & \second{1.70} & \best{6.69} & \best{0.34} & \second{0.33} \\
   & MiDashengLM-Gen & \best{0.98} & \second{6.78} & \second{0.40} & \best{0.34} \\
  \midrule
  \multirow{5}{*}{SMA} & TangoFlux & 11.40 & 50.58 & 2.05 & 0.29 \\
   & MusicGen & 20.18 & 69.19 & 3.41 & 0.11 \\
   & Qwen3-TTS & 16.53 & 83.01 & 2.45 & 0.10 \\
   & Dasheng AudioGen & \second{2.17} & \best{17.75} & \best{0.63} & \best{0.38} \\
   & MiDashengLM-Gen & \best{1.88} & \second{20.60} & \second{0.85} & \second{0.35} \\
  \bottomrule
  \end{tabular}
  \end{table}

\subsection{Ablation Experiments}
\label{ssec:ablation}

\noindent We ablate two core designs of MiDashengLM-Gen: the audio-text alignment stage and the DiT architecture for the flow matching decoder.

\noindent{\textbf{Importance of Audio-Text Alignment.}}
To validate the necessity of the audio-text alignment stage (Section~\ref{ssec:audio_text_alignment}), we compare our full model against a variant that skips alignment entirely: the audio adapter (projection layer) is randomly initialized and the LLM retains its original Qwen3 pre-trained weights without alignment training, while the DashengTokenizer encoder keeps its pre-trained weights. The generation training procedure remains identical in both settings.
Table~\ref{tab:alignment_ablation} shows that removing audio-text alignment degrades performance across all benchmarks. FAD increases by 47--88\% on audio generation tasks (e.g., 5.01$\to$9.39 on AudioCaps), indicating weakened audio quality and text relevance. The impact on speech generation is particularly severe: Seed-TTS English WER rises from 2.79\% to 12.17\% (4.4$\times$), Chinese CER from 3.87\% to 14.17\% (3.7$\times$), and mean multilingual WER from 7.68\% to 31.73\% (4.1$\times$). This disproportionate degradation in speech intelligibility suggests that audio-text alignment is especially critical for speech generation, where the LLM must map text semantics precisely to phonetic content. Without a pre-aligned embedding space, the generation stage cannot effectively leverage the LLM's language understanding for intelligible speech synthesis.

\begin{table}[t]
  \centering
  \caption{Ablation on audio-text alignment. ``w/o Align'' uses randomly initialized audio adapter and original Qwen3 LLM weights without alignment pre-training. MECAT Single/Mixed report weighted averages across their respective categories; Multilingual reports the mean across 9 languages.}
  \label{tab:alignment_ablation}
  \small
  \begin{tabular}{llcc}
  \toprule
  Benchmark & Metric & w/o Align & MiDashengLM-Gen \\
  \midrule
  \multirow{2}{*}{AudioCaps} & FAD $\downarrow$ & 9.39 & \best{5.01} \\
   & CLAP $\uparrow$ & 0.28 & \best{0.36} \\
  \midrule
  \multirow{2}{*}{MusicCaps} & FAD $\downarrow$ & 4.06 & \best{2.28} \\
   & CLAP $\uparrow$ & 0.30 & \best{0.33} \\
  \midrule
  \multirow{2}{*}{MECAT Single} & FAD $\downarrow$ & 3.18 & \best{1.86} \\
   & CLAP $\uparrow$ & 0.32 & \best{0.33} \\
  \midrule
  \multirow{2}{*}{MECAT Mixed} & FAD $\downarrow$ & 2.46 & \best{1.35} \\
   & CLAP $\uparrow$ & 0.32 & \best{0.35} \\
  \midrule
  \multirow{2}{*}{Seed-TTS} & EN WER(\%) $\downarrow$ & 12.17 & \best{2.79} \\
   & ZH CER(\%) $\downarrow$ & 14.17 & \best{3.87} \\
  \midrule
  Multilingual & Mean WER(\%) $\downarrow$ & 31.73 & \best{7.68} \\
  \bottomrule
  \end{tabular}
\end{table}

\begin{figure}[t]
  \centering
  \includegraphics[width=0.8\columnwidth]{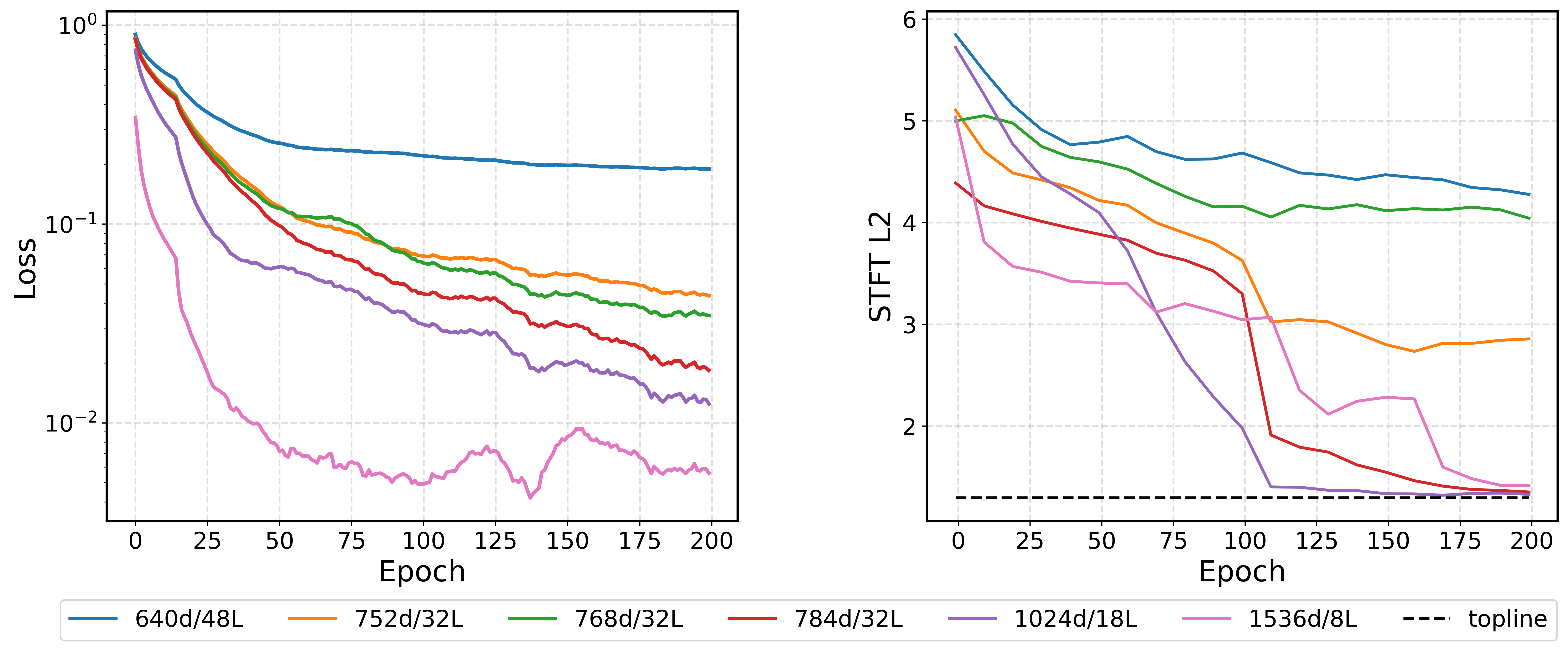}
  \caption{Training loss and STFT L2 distance curves for different DiT width/depth configurations. Topline denotes DashengTokenizer encode-to-decode reconstruction.}
  \label{fig:metrics_curves}
\end{figure}

\noindent{\textbf{Scaling DiT Width to Match Audio Latent Dimensionality.}}
Recent work on diffusion transformers has shown that the model width must match the token dimensionality of the latent space, because noise injection in diffusion training expands the data manifold to a full-rank distribution~\cite{zheng2025diffusion}. We investigate whether this principle extends to audio generation with MiDashengLM-Gen. To this end, we conduct single-sample overfit experiments with varying DiT width and depth configurations under a fixed parameter budget. Figure~\ref{fig:metrics_curves} shows the training loss and STFT L2 distance curves for our final model configuration (768-dim audio latent + Qwen3-1.7B), with the DashengTokenizer encode-to-decode reconstruction as the topline. Table~\ref{tab:width_convergence} further validates the convergence pattern across three additional model scales, with the DiT total parameters fixed across all configurations by trading off width and depth.

\begin{table}[t]
  \centering
  \caption{Convergence analysis of DiT width across different model scales. \cmark denotes successful convergence.}
  \label{tab:width_convergence}
  \small
  \begin{tabular}{lccc}
  \toprule
  Setting & DiT Width & Train Loss $\downarrow$ & Convergence \\
  \midrule
  \multirow{4}{*}{768-dim + Qwen3 0.6B}
    & 752 & 0.0422 & \xmark \\
    & 768 & 0.0343 & \xmark \\
    & 784 & 0.0171 & \cmark \\
    & 1024 & 0.0110 & \cmark \\
  \midrule
  \multirow{4}{*}{1280-dim + Qwen3 0.6B}
    & 1248 & 0.0354 & \xmark \\
    & 1280 & 0.0144 & \cmark \\
    & 1312 & 0.0084 & \cmark \\
    & 1536 & 0.0064 & \cmark \\
  \midrule
  \multirow{4}{*}{1280-dim + Qwen3 4B}
    & 1248 & 0.0365 & \xmark \\
    & 1280 & 0.0126 & \cmark \\
    & 1312 & 0.0081 & \cmark \\
    & 1536 & 0.0063 & \cmark \\
  \bottomrule
  \end{tabular}
\end{table}

Figure~\ref{fig:metrics_curves} reveals that the DiT width must strictly exceed the audio latent dimensionality for successful convergence. When width falls below or equals this threshold (768 for 768-dim audio), the model fails to converge (\xmark) with an STFT L2 distance of 3.89, well above the topline of 1.30. Only once width exceeds the latent dimensionality does training converge successfully (\cmark). Table~\ref{tab:width_convergence} confirms that this pattern generalizes across different LLM scales (0.6B and 4B) and audio latent dimensionalities (768 and 1280), mirroring the theoretical finding by Zheng et al.~\cite{zheng2025diffusion} in image diffusion.

\section{Discussion and Conclusion}
\label{sec:discussion}

We presented MiDashengLM-Gen, an autoregressive approach that combines a pre-trained LLM backbone with per-token flow matching for variable-length mixed-audio scene generation.
The key insight is that high-dimensional semantic-acoustic latents are viable for LLM-conditioned generation when the decoder architecture is properly scaled — specifically, the DiT width must exceed the audio latent dimensionality for convergence.

Comprehensive evaluations demonstrate that MiDashengLM-Gen substantially improves speech intelligibility over prior unified models (2.79\% vs.\ 12.15\%~\cite{mei2026dashengaudiogen} on Seed-TTS English), approaching dedicated TTS systems (1.24\%), while maintaining competitive mixed-audio scene generation and supporting multilingual output.
Ablation studies reveal two critical design requirements: audio-text alignment pre-training is essential for bridging the modality gap (removing it degrades speech WER by $4\times$), and the DiT decoder width must strictly exceed the audio latent dimensionality, consistent with theoretical findings in image diffusion~\cite{zheng2025diffusion}.

MiDashengLM-Gen has several limitations.
Variable-length generation is bounded by the training data distribution (1--20 seconds); generating coherent audio beyond this range remains open.
Speech intelligibility still trails dedicated TTS systems by a factor of 2.2$\times$ on English, with larger gaps for low-resource languages where TTS training data is scarce.
The model supports only coarse speaker-style control, without voice cloning or explicit speaker identity conditioning.

Future work will explore scaling to longer durations, improved low-resource language performance, voice cloning, and finer-grained controllability such as audio editing and explicit temporal control.

\section{Acknowledgement}
\label{sec:acknowledgement}

This work makes use of the ACAVCaps~\cite{acavcaps}, Emilia~\cite{he2024emilia}, LibriTTS~\cite{LibriTTS}, LJSpeech~\cite{ljspeech17}, AISHELL-3~\cite{shi2020aishell3}, and WenetSpeech4TTS~\cite{ma2024wenetspeech4tts} datasets for training. For evaluation, we use the AudioCaps~\cite{kim2019audiocaps}, MusicCaps~\cite{agostinelli2023musiclm}, Seed-TTS~\cite{seedtts}, MiniMax-Speech~\cite{zhang2025minimaxspeech}, CV3-Eval~\cite{du2025cosyvoice3}, and MECAT~\cite{nuu2025mecat} benchmarks. The authors confirm that the use of the ACAVCaps, Emilia, AISHELL-3, and WenetSpeech4TTS datasets is strictly limited to academic research purposes and does not involve any commercial activities. All datasets are used in compliance with their respective licensing agreements and original citations.

% \clearpage

\bibliographystyle{unsrt}
\bibliography{references}

\end{document}